\documentclass[aps, pra, twocolumn, amsmath, amssymb, superscriptaddress, nofootinbib]{revtex4-1}

\makeatletter
\def\p@subsection{}
\def\p@subsubsection{}
\makeatother

\usepackage[utf8]{inputenc}
\usepackage{graphicx}
\usepackage{units}
\usepackage{color}
\usepackage[pdftex, colorlinks=true, linkcolor=myblue, citecolor=myblue,
urlcolor=myblue]{hyperref}
\usepackage{times}
\usepackage{comment}
\usepackage{graphicx}
\usepackage{amsmath, calc}
\usepackage{mathrsfs}
\usepackage{amsfonts}
\usepackage{amssymb}
\usepackage{bm}
\usepackage{bbm}
\usepackage{color}
\usepackage{array}
\usepackage{units}
\usepackage{tabstackengine}
\stackMath

\definecolor{myblue}{rgb}{0,0,1}
\definecolor{myred}{rgb}{1,0,0}

\begin{document}


\title{Directionality between driven-dissipative resonators}


\author{C. A. Downing}
\email{c.a.downing@exeter.ac.uk} 
\affiliation{Department of Physics and Astronomy, University of Exeter, Exeter EX4 4QL, United Kingdom}

\author{T.~J.~Sturges}
\affiliation{Institute of Theoretical Solid State Physics, Karlsruhe Institute of Technology (KIT), D-76131 Karlsruhe, Germany}


\date{\today}


\begin{abstract}
\noindent
\\
 \textbf{Abstract}\\
The notion of nonreciprocity, in essence when going forwards is different from going backwards, emerges in all branches of physics from cosmology to electromagnetism. Intriguingly, the breakdown of reciprocity is typically associated with extraordinary phenomena, which may be readily capitalized on in the design of (for example) nontrivial electromagnetic devices when Lorentz reciprocity is broken. However, in order to enable the exploitation of nonreciprocal-like effects in the next generation of quantum technologies, basic quantum optical theories are required. Here we present a versatile model describing a pair of driven-dissipative quantum resonators, where the relative phase difference between the coherent and incoherent couplings induces an asymmetry. The interplay between the diverse dissipative landscape -- which encompasses both intrinsic losses and dissipative couplings -- and the coherent interactions leads to some remarkable consequences including highly directional (or even one-way) energy transport. Our work proffers the tantalizing prospect of observing dissipation-induced quantum directionality in areas like photonics or cavity magnonics (spin waves), which may aid the design of unconventional nanoscopic devices.
\end{abstract}


\maketitle



\noindent \textbf{Introduction}\\
In perhaps one of the most profound conversations in Confucianism, Tsze-kung asks ``Is there one word which may serve as a rule of practice for all one's life?'' The Master replies ``Is not \textit{shu} [reciprocity] such a word? What you do not want done to yourself, do not do to others''~\cite{Legge1971}.

In physics, reciprocity is a famous hallmark across the entire discipline, from the equal and opposite quality of Newton's third law of motion~\cite{Feynman1989} to the Lorentz reciprocity in electromagnetism, which guarantees the same response when the source and receiver are interchanged~\cite{Caloz2018}. Introducing various types of asymmetries can lead to a treasure trove of curiosities, which in optics precipitated the exciting subfields of chiral plasmonics~\cite{Hentschel2017, Andrews2018} and chiral quantum optics~\cite{Lodahl2017, Downing2019, Sanchez2019}. Within device physics, it has already been shown that exploiting nonreciprocity can give rise to practical applications like high quality factor, large bandwidth devices~\cite{Tsakmakidis2017} (which are predicated upon the induced asymmetric transport properties), optical isolators~\cite{Jalas2013} (where the governing scattering matrix is inherently asymmetric), and even magnetic diodes~\cite{Prat2018} (where it was demonstrated that the magnetic coupling between two coils above a conductor - moving with constant velocity - may become asymmetric, leading to a diode for magnetic fields).

\begin{figure}[tb]
 \includegraphics[width=1.0\linewidth]{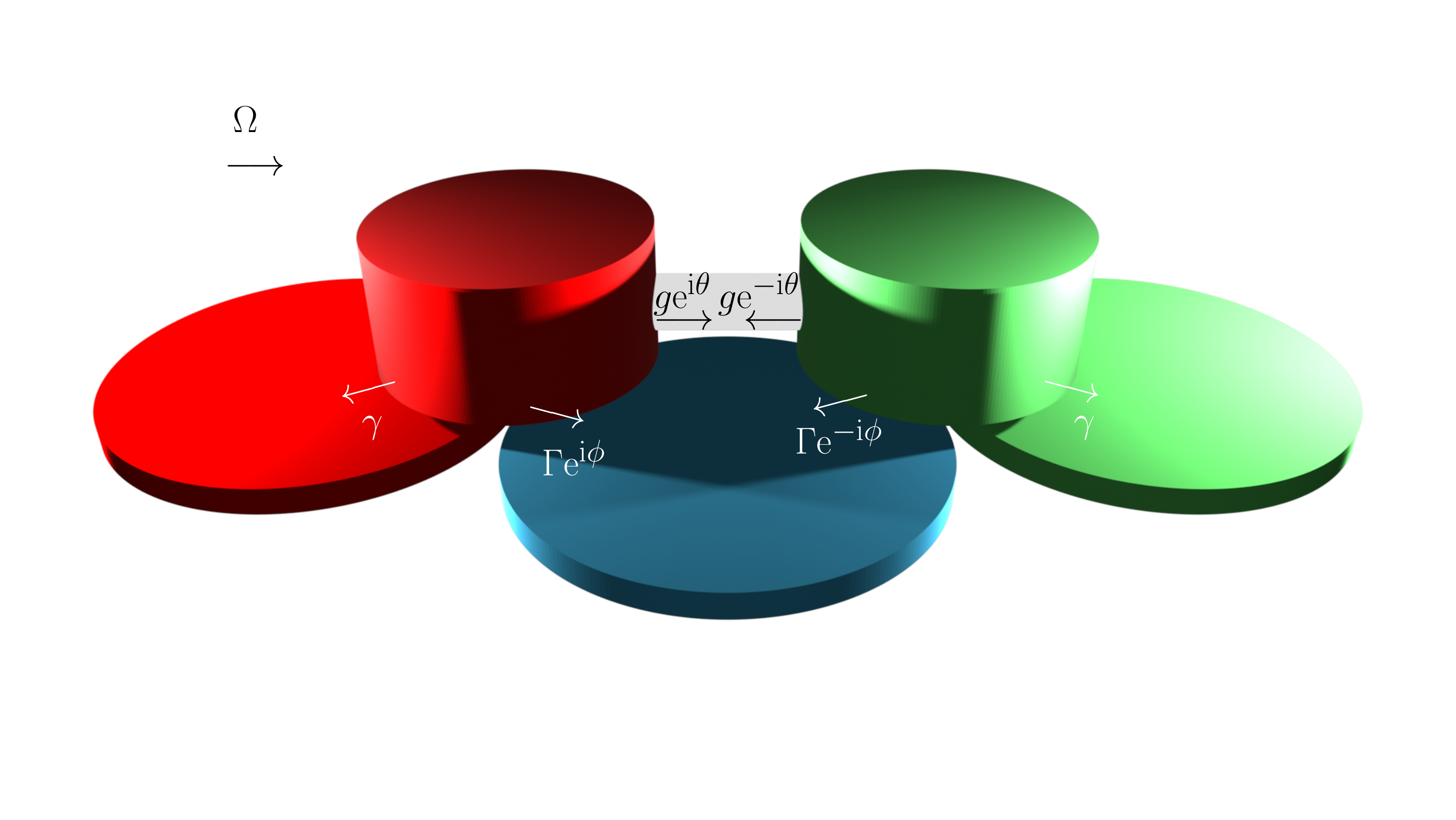}
 \caption{ \textbf{A sketch of a pair of driven-dissipative resonators.} The first resonator (red pillar) is driven by a laser with amplitude $\Omega$, while the second resonator (green pillar) is undriven [cf. Eq.~\eqref{eq:Hammy}]. The coherent coupling (of magnitude $g$ and phase $\theta$) between the resonators is represented by the gray rod, while the dissipative coupling (of magnitude $\Gamma$ and phase $\phi$) is mediated by the common bath (blue disk) [cf. Eq.~\eqref{eqapp:massdsdsdter}]. The individual loss $\gamma$ of each resonator is associated with the red and green disks.}
 \label{mycartoon}
\end{figure}

Here we investigate a basic quantum optical model, namely a pair of driven-dissipative resonators, with a view to inducing asymmetric behaviour~\cite{Metelmann2015, Metelmann2017,Kamal2017, Downing2020}. We consider two oscillators, which are in general coupled both coherently and incoherently, where the first resonator is additionally coherently driven by a laser as sketched in Fig.~\ref{mycartoon}. The admixture between the coherent and incoherent couplings, which are in general complex quantities, has profound consequences for directionality in the system. There are four principle coupling regimes of our model: (i) \textit{coherent} coupling, where the direct hopping between the resonators dominates (gray rod in the figure), (ii) \textit{dissipative} coupling, where incoherent coupling between the resonators is of primary importance (as mediated by their common heat bath, the blue disk in the figure), (iii) \textit{unidirectional} coupling, where the coupling between the resonators is completely one-way, and (iv) \textit{asymmetric} coupling, where the mixture of coherent and incoherent coupling leads to asymmetries in the interactions between the resonator pair [such that this case generalizes the more extreme limiting case (iii)]. In what follows, we provide a simple analysis of the population dynamics in these regimes.
\\

\begin{figure*}[tb]
 \includegraphics[width=1.0\linewidth]{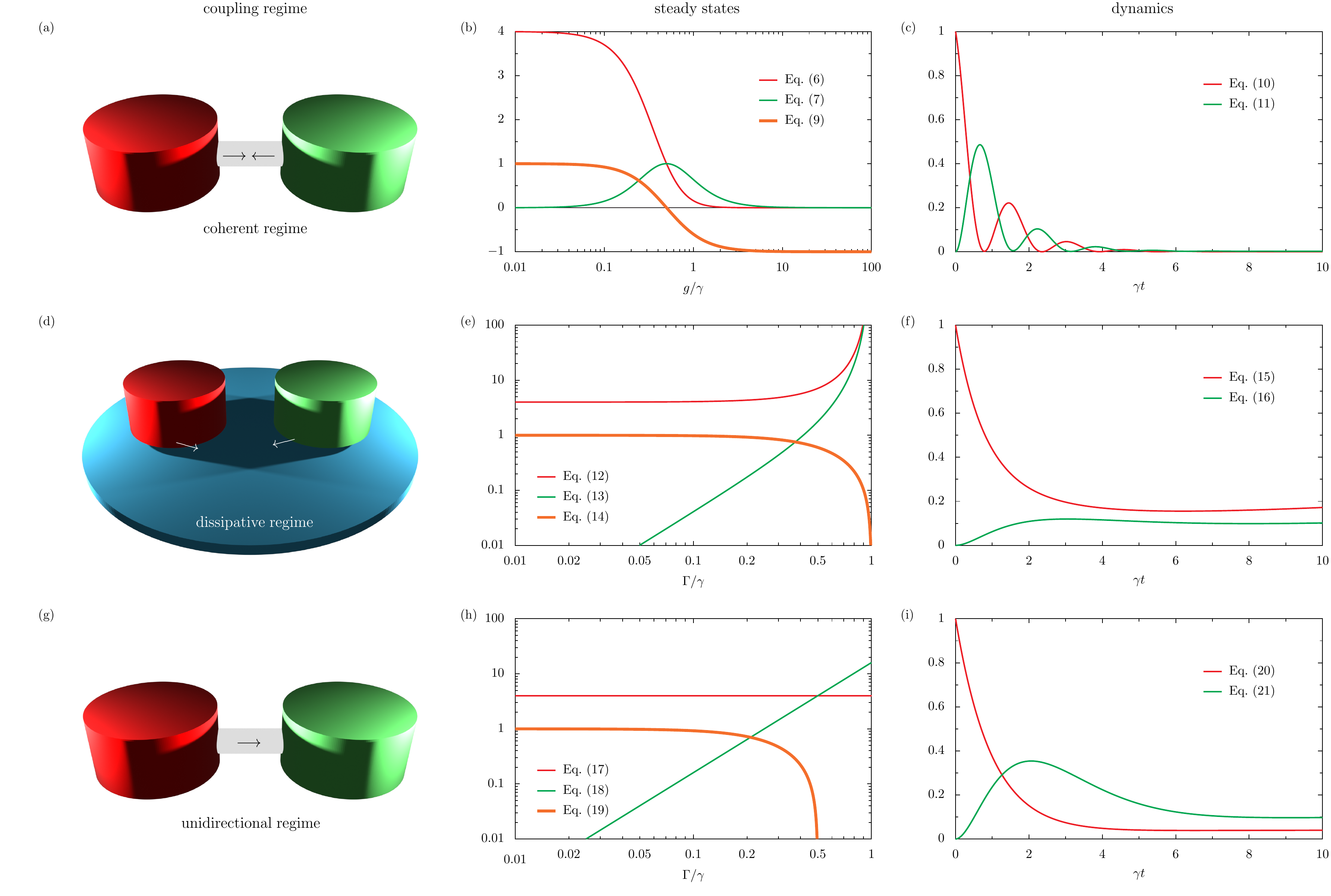}
 \caption{ \textbf{The populations of a pair of driven-dissipative resonators.} First column: sketches of the coherent, dissipative and unidirectional coupling regimes. Second column: the scaled steady state populations $\lim_{t \to \infty} \langle b_n^{\dagger} b_n \rangle \times \left( \gamma / \Omega \right)^2$ of the pair (red and green lines), as a function of the coherent coupling strength $g$ or the dissipative coupling strength $\Gamma$. The population imbalance $\Delta$ is shown as thick orange lines [cf. Eq.~\eqref{eqapp:dssdfdfdfdfdsaysdshr}]. Third column: the dynamic populations $\langle b_n^{\dagger} b_n \rangle$ of the pair, as a function of time $t$ (in units of the inverse loss rate $\gamma^{-1}$). In the figure, the first resonator is driven with an amplitude $\Omega = \gamma/10$, in panel (c) $g = 2\gamma$ and in panels (f) and (i) $\Gamma = (4/5)\gamma$. }
 \label{poppp}
\end{figure*}


\noindent \textbf{Model}\\
The driven coupled oscillators model represented in Fig.~\ref{mycartoon} may be described by the Hamiltonian $\hat{H}$ (with $\hbar = 1$)
\begin{align}
\label{eq:Hammy}
 \hat{H} =&~\omega_\Delta \left( b_1^{\dagger} b_1 +  b_2^{\dagger} b_2 \right) + \Omega \left( b_1^{\dagger} + b_1 \right) \nonumber \\
 &+  g \mathrm{e}^{\mathrm{i} \theta} b_1^{\dagger} b_2 +  g \mathrm{e}^{-\mathrm{i} \theta} b_2^{\dagger} b_1,
\end{align}
where the $n$-th oscillator sustains bosonic excitations created (destroyed) by the operator $b_n^\dagger$ ($b_n$). The coherent oscillator-oscillator coupling is of strength $g \ge 0$ and phase $\theta$. The first oscillator is driven by a laser of amplitude $\Omega$, and the detunings $\omega_\Delta$ arise in the chosen rotating reference frame of the laser~\cite{SuppInfo}.

We include dissipation in the model through a quantum master equation, which describes the time evolution of the density matrix $\rho$ of the system via~\cite{SuppInfo, Gardiner2014, Dung2002} 
\begin{align}
\label{eqapp:massdsdsdter}
 \partial_t \rho =&~~\mathrm{i} [ \rho, \hat{H} ] \nonumber \\
&+ \sum_{n=1,2} \frac{\gamma}{2} \left( 2 b_n \rho b_n^{\dagger} -  b_n^{\dagger} b_n \rho - \rho b_n^{\dagger} b_n \right) \nonumber \\
&+  \frac{\Gamma \mathrm{e}^{\mathrm{i}\phi}}{2} \left( 2 b_2 \rho b_1^{\dagger} -  b_1^{\dagger} b_2 \rho - \rho b_1^{\dagger} b_2 \right) \nonumber \\
&+  \frac{\Gamma \mathrm{e}^{-\mathrm{i}\phi}}{2} \left( 2 b_1 \rho b_2^{\dagger} -  b_2^{\dagger} b_1 \rho - \rho b_2^{\dagger} b_1 \right). 
\end{align}
The first line of Eq.~\eqref{eqapp:massdsdsdter} is the von Neumann equation, describing the unitary evolution of the closed system as governed by the Hamiltonian $\hat{H}$ of Eq.~\eqref{eq:Hammy}. The non-unitary evolution is captured by the three lower lines of Eq.~\eqref{eqapp:massdsdsdter}, which are written as Lindblad terms and describe the open quantum system sketched in Fig.~\ref{mycartoon}. In particular, the second line on the right-hand-side of Eq.~\eqref{eqapp:massdsdsdter} accounts for the intrinsic loss $\gamma$ of each oscillator. The third and fourth lines of Eq.~\eqref{eqapp:massdsdsdter} track the dissipative (incoherent) coupling between the pair of oscillators due to their shared heat bath, which can generally be regarded as a complex quantity of magnitude $\Gamma$ and phase $\phi$ (subject to the condition $0 \le \Gamma \le \gamma$).

The first moments $\langle b_n \rangle$ of the system described by Eq.~\eqref{eq:Hammy} and Eq.~\eqref{eqapp:massdsdsdter} may be found by the following pair of coupled first-order equations~\cite{SuppInfo}
\begin{equation}
\label{eq:Hamdfdfdfsdfsdfmy}
\mathrm{i} \partial_t 
\begin{pmatrix}
\langle b_1 \rangle \\
\langle b_2 \rangle
\end{pmatrix} = 
\begin{pmatrix}
\omega_\Delta - \mathrm{i} \frac{\gamma}{2} & G_-\\
G_+^\ast & \omega_\Delta - \mathrm{i} \frac{\gamma}{2}
\end{pmatrix}
\begin{pmatrix}
\langle b_1 \rangle \\
\langle b_2 \rangle
\end{pmatrix}
+
\begin{pmatrix}
\Omega\\
0
\end{pmatrix},
\end{equation}
where we have introduced two generalized coupling constants $G_{+}$ and $G_{-}$, hereby defined as
\begin{equation}
\label{eq:sdf}
G_{\pm} = g \mathrm{e}^{\mathrm{i} \theta} \pm \tfrac{1}{2} \Gamma\mathrm{e}^{\mathrm{i} \phi},
\end{equation}
which accounts for the admixture between the competing coherent and dissipative couplings, including their magnitudes and phases. Most notably, this analysis reveals that the oscillator-oscillator coupling may be completely unidirectional, as was first noticed in the celebrated works on cascaded quantum systems~\cite{Gardiner1993, Carmichael1993}. When $G_- = 0$ in Eq.~\eqref{eq:Hamdfdfdfsdfsdfmy} one-way coupling in the rightwards direction ($\rightarrow$) arises, and likewise when $G_+^\ast = 0$ in Eq.~\eqref{eq:Hamdfdfdfsdfsdfmy} the coupling is completely leftwards ($\leftarrow$).  Amongst the entire space of possible couplings, these two special circumstances occur when the following conditions on the coupling magnitudes and relative phases hold [cf. Eq.~\eqref{eq:sdf}]
\begin{equation}
\label{eq:Hamsdfsdfmy}
 \Gamma = 2 g, 
 \quad\quad\quad\quad
  \theta - \phi = \begin{cases}
 \tfrac{\pi}{2}  & \left( \rightarrow \right) \\
  \tfrac{3\pi}{2} & \left( \leftarrow  \right)
\end{cases}.
\end{equation}
Away from these twin conditions for unidirectionality, the overall coupling is generally asymmetric. Wonderfully, the basic theoretical model encapsulated by Eq.~\eqref{eq:Hammy} and Eq.~\eqref{eqapp:massdsdsdter} may be realized in an eclectic range of systems. For example, in spin-photon systems such as cavity magnons with coherent and dissipative couplings~\cite{Wang2020, Harder2021}, in circuit-QED setups utilizing superconducting qubits~\cite{Wendin2017, Blais2021}, in plasmonic epsilon-near-zero waveguides~\cite{Issah2021}, in metallic nanoparticle architectures exploiting plasmonic responses~\cite{Sturges2020}, and with coupled cavity-based photonic devices~\cite{Metelmann2015, Azcona2021}. In what follows we keep our discussion of the model general, keeping in mind specializations are readily obtainable experimentally.
\\


\noindent \textbf{Coherent coupling}\\
Let us first consider the simplest case of purely coherent coupling between the resonators (so that $\Gamma = 0$), as is represented in the sketch of Fig.~\ref{poppp}~(a). We are interested in the populations $\langle b_n^{\dagger} b_n \rangle$ of the resonators, which may be calculated from the second moment analogue of Eq.~\eqref{eq:Hamdfdfdfsdfsdfmy}, as discussed in Ref.~\cite{SuppInfo}.

At long time scales, the competition between the driving and dissipation leads to a well-defined steady state, described by the analytic expressions~\cite{SuppInfo}
\begin{equation}
\label{eqapp:dsayhr}
\lim_{t \to \infty} \langle b_1^{\dagger} b_1 \rangle = \left( \frac{2 \gamma \Omega}{\gamma^2 + 4 g^2}\right)^2,
\end{equation}
\begin{equation}
\label{eqapp:dsaysdshr}
\lim_{t \to \infty} \langle b_2^{\dagger} b_2 \rangle = \left( \frac{4 g \Omega}{\gamma^2 + 4 g^2}\right)^2.
\end{equation}
These populations, scaled by $\left( \gamma / \Omega \right)^2$, are plotted as a function of the coherent coupling strength $g$ for the first resonator (red line) and second resonator (green line) in Fig.~\ref{poppp}~(b). Clearly, since only the first resonator is driven, its steady state population is bounded by its maximum of $\left( 2 \Omega / \gamma \right)^2$ when $g \ll \gamma$, and decreases to $ \{   \gamma \Omega  / \left( 2 g^2 \right) \}^2$ when $g \gg \gamma$. In these limits, the second (and undriven) resonator population is zero when $g \ll \gamma$ and $\left( \Omega / g \right)^2$ when $g \gg \gamma$, its maximum population of $\left( \Omega / \gamma \right)^2$ is instead met when $g = \gamma/2$. Hence the population imbalance between the resonators is a useful quantity to describe the system, in the steady state it reads
\begin{equation}
\label{eqapp:dssdfdfdfdfdsaysdshr}
\Delta = \lim_{t \to \infty}  \frac{ \langle b_1^{\dagger} b_1 \rangle - \langle b_2^{\dagger} b_2 \rangle }{ \langle b_1^{\dagger} b_1 \rangle + \langle b_2^{\dagger} b_2 \rangle }.
\end{equation}
In this coherent coupling regime, $\Delta$ is a sign-changing quantity, which is explicitly given by
\begin{equation}
\label{eqapp:dssdfdsaysdshr}
\Delta = 1 - \frac{8 g^2}{\gamma^2 + 4 g^2},
\end{equation}
which observes the bounds of $-1 \le \Delta \le 1$, as displayed with the thick orange line in Fig.~\ref{poppp}~(b). The critical point of a completely balanced populations across the resonators $\Delta = 0$ is reached when $g = \gamma /2$, and above this coupling strength the second resonator has a larger steady state population despite being undriven.

The full dynamic populations $\langle b_n^{\dagger} b_n \rangle$ of the coupled resonators are given by the exact equations~\cite{SuppInfo}
\begin{widetext}
\begin{align}
\label{eqapp:dssdfsfayhr}
 \langle b_1^{\dagger} b_1 \rangle =& \left( \frac{2 \gamma \Omega}{\gamma^2 + 4 g^2}\right)^2 + 2 \left( \frac{2 \gamma \Omega}{\gamma^2 + 4 g^2}\right)^2 \biggl\{ 2 g \sin \left( g t \right) - \gamma \cos \left( g t \right) \biggl\} \mathrm{e}^{-\frac{\gamma t}{2}}  \\
 &+ \biggl\{ \Bigr[ \gamma^2 + 4 g^2 \Bigr] \Bigr[ \gamma^2 + 4 g^2 + 4 \Omega^2 \Bigr] + \Bigr[  \left( \gamma^2 + 4 g^2 \right)^2 + 4 \Omega^2 \left( \gamma^2 - 4 g^2 \right) \Bigr] \cos \left( 2 g t\right) - 16 g \gamma \Omega^2 \sin \left( 2 g t\right)  \biggl\} \frac{\mathrm{e}^{-\gamma t}}{2 \left( \gamma^2 + 4 g^2 \right)^2}, \nonumber \\
  \langle b_2^{\dagger} b_2 \rangle =& \left( \frac{4 g \Omega}{\gamma^2 + 4 g^2}\right)^2 - \left( \frac{4 g \Omega}{\gamma^2 + 4 g^2}\right)^2 \biggl\{ 2 g \cos \left( g t \right) + \gamma \sin \left( g t \right) \biggl\} \mathrm{e}^{-\frac{\gamma t}{2}}  \\
 &+ \biggl\{ \Bigr[ \gamma^2 + 4 g^2 \Bigr] \Bigr[ \gamma^2 + 4 g^2 + 4 \Omega^2 \Bigr] - \Bigr[  \left( \gamma^2 + 4 g^2 \right)^2 + 4 \Omega^2 \left( \gamma^2 - 4 g^2 \right) \Bigr] \cos \left( 2 g t\right) + 16 g \gamma \Omega^2 \sin \left( 2 g t\right)  \biggl\} \frac{\mathrm{e}^{-\gamma t}}{2 \left( \gamma^2 + 4 g^2 \right)^2}, \nonumber
\end{align}
\end{widetext}
which are plotted in Fig.~\ref{poppp}~(c) as a function of time (for the example case of $g = 2 \gamma$). The characteristic damped Rabi oscillations are shown, and the two different time constants appearing in the above expressions ($1/\gamma$ and $2/\gamma$) is characteristic of coherently driven systems. This sets out the stall for the most well known coupling regime.
\\


\noindent \textbf{Dissipative coupling}\\
When the coherent coupling is negligible (that is, $g = 0$), the system is in the dissipative coupling regime. This setup is sketched in Fig.~\ref{poppp}~(d), where the blue disk represents the common heat bath enabling the incoherent coupling. In the steady state, the resonator populations are described by the simple forms~\cite{SuppInfo}
\begin{equation}
\label{eqapp:dsasdsdyhr}
\lim_{t \to \infty} \langle b_1^{\dagger} b_1 \rangle = \left( \frac{2 \gamma \Omega}{\gamma^2 - \Gamma^2}\right)^2,
\end{equation}
\begin{equation}
\label{eqapp:dssdsdaysdshr}
\lim_{t \to \infty} \langle b_2^{\dagger} b_2 \rangle = \left( \frac{2 \Gamma \Omega}{\gamma^2 - \Gamma^2}\right)^2.
\end{equation}
In the weak dissipative coupling limit $\Gamma \ll \gamma$ of course only the first, driven resonator is populated, with $\left( 2 \Omega / \gamma \right)^2$. In the opposing strong dissipative coupling $\Gamma \to \gamma$ limit the bosonic nature of the resonators becomes readily apparent, since both populations tend towards infinity, as is shown in in Fig.~\ref{poppp}~(e). The population imbalance [cf. Eq.~\eqref{eqapp:dssdfdfdfdfdsaysdshr}] is always singled-signed and reads
\begin{equation}
\label{eqapp:dssdfddgfsaysdshr}
\Delta = 1 - \frac{2 \Gamma^2}{\Gamma^2 + \gamma^2},
\end{equation}
which exposes the necessarily non-negative bounds of $0 \le \Delta \le 1$, as show by the thick orange line in Fig.~\ref{poppp}~(e). This in stark contrast to the coherent coupling regime, which allows for positive and negative imbalances [cf. panel (b)].

The time-dependent populations are described by the following expressions~\cite{SuppInfo}
\begin{widetext}
\begin{align}
\label{eqapp:dssdfdfdfsfayhr}
 \langle b_1^{\dagger} b_1 \rangle =& \left( \frac{2 \gamma \Omega}{\gamma^2 - \Gamma^2}\right)^2 + \left(  \cosh \left( \Gamma t \right) + 1 + \frac{4\Omega^2}{\gamma^2-\Gamma^2} \right) \frac{\mathrm{e}^{-\gamma t}}{2} + \Omega^2 \biggl\{ \frac{\mathrm{e}^{- \left( \gamma + \Gamma \right) t}}{\left( \gamma + \Gamma \right)^2} + \frac{\mathrm{e}^{- \left( \gamma - \Gamma \right) t}}{\left( \gamma - \Gamma \right)^2} \biggl\} \nonumber \\
& - \frac{4 \gamma \Omega^2}{ \gamma^2 - \Gamma^2 } \biggl\{ \frac{\mathrm{e}^{- \frac{\left( \gamma + \Gamma \right) t}{2}}}{ \gamma + \Gamma} + \frac{\mathrm{e}^{- \frac{ \left( \gamma - \Gamma \right) t}{2}}}{\gamma - \Gamma} \biggl\}, \\
 \langle b_2^{\dagger} b_2 \rangle =& \left( \frac{2 \Gamma \Omega}{\gamma^2 - \Gamma^2}\right)^2 + \left(  \cosh \left( \Gamma t \right) - 1 - \frac{4\Omega^2}{\gamma^2-\Gamma^2} \right) \frac{\mathrm{e}^{-\gamma t}}{2} + \Omega^2 \biggl\{ \frac{\mathrm{e}^{- \left( \gamma + \Gamma \right) t}}{\left( \gamma + \Gamma \right)^2} + \frac{\mathrm{e}^{- \left( \gamma - \Gamma \right) t}}{\left( \gamma - \Gamma \right)^2} \biggl\} \nonumber \\
& + \frac{4 \Gamma \Omega^2}{ \gamma^2 - \Gamma^2 } \biggl\{ \frac{\mathrm{e}^{- \frac{\left( \gamma + \Gamma \right) t}{2}}}{ \gamma + \Gamma} - \frac{\mathrm{e}^{- \frac{ \left( \gamma - \Gamma \right) t}{2}}}{\gamma - \Gamma} \biggl\}, 
\end{align}
\end{widetext}
as displayed in Fig.~\ref{poppp}~(f) for the example case of reasonably strong dissipative coupling $\Gamma = (4/5)\gamma$. A hallmark of this coupling regime is the lack of any Rabi oscillations due to the absence of any coherent coupling, and the supremacy in population of the first resonator for any value of the dissipative coupling strength $\Gamma$, as suggested by Eq.~\eqref{eqapp:dssdfddgfsaysdshr}.
\\


\noindent \textbf{Unidirectional coupling}\\
The final special case of coupling that we shall consider is that of unidirectional coupling, and in particular when the conditions of Eq.~\eqref{eq:Hamsdfsdfmy} are met for the rightwards ($\rightarrow$) direction only, as represented by the picture in Fig.~\ref{poppp}~(g). The lack of backaction ensures that the first resonator population is coupling independent (in this regime, the dissipative coupling strength $\Gamma = 2g$ is fixed) while the second resonator population is enhanced due to the one-way nature of the interaction. The steady state results are simply~\cite{SuppInfo}
\begin{equation}
\label{eqapp:dsxvcjhdfhdayhr}
\lim_{t \to \infty} \langle b_1^{\dagger} b_1 \rangle = \left( \frac{2 \Omega}{\gamma^2}\right)^2,
\end{equation}
\begin{equation}
\label{eqapdasdp:dssfsaysdshr}
\lim_{t \to \infty} \langle b_2^{\dagger} b_2 \rangle = \left( \frac{4 \Gamma \Omega}{\gamma^2}\right)^2,
\end{equation}
which are plotted in Fig.~\ref{poppp}~(h) as a function of $\Gamma$. As must be the case, the first resonator population (red line) is exactly that of a single driven-dissipative oscillator~\cite{SuppInfo}, while the second resonator resonator presents a quadratic scaling with the dissipative coupling $\Gamma$. Therefore, the population imbalance [cf. Eq.~\eqref{eqapp:dssdfdfdfdfdsaysdshr}] is given by
\begin{equation}
\label{eqapsfap:dafssdfdsaysdshr}
\Delta = 1 - \frac{8 \Gamma^2}{\gamma^2 + 4 \Gamma^2},
\end{equation}
as plotted as the thick orange line in Fig.~\ref{poppp}~(h). Unlike in the dissipative coupling regime, this population imbalance is a sign-changing quantity, being bounded by $-3/5 \le \Delta \le 1$. The critical point of $\Delta = 0$ is reached when $\Gamma = \gamma/2$, such that above this dissipative coupling strength the undriven second resonator is more highly populated that the driven and first resonator.

\begin{figure}[tb]
 \includegraphics[width=1.0\linewidth]{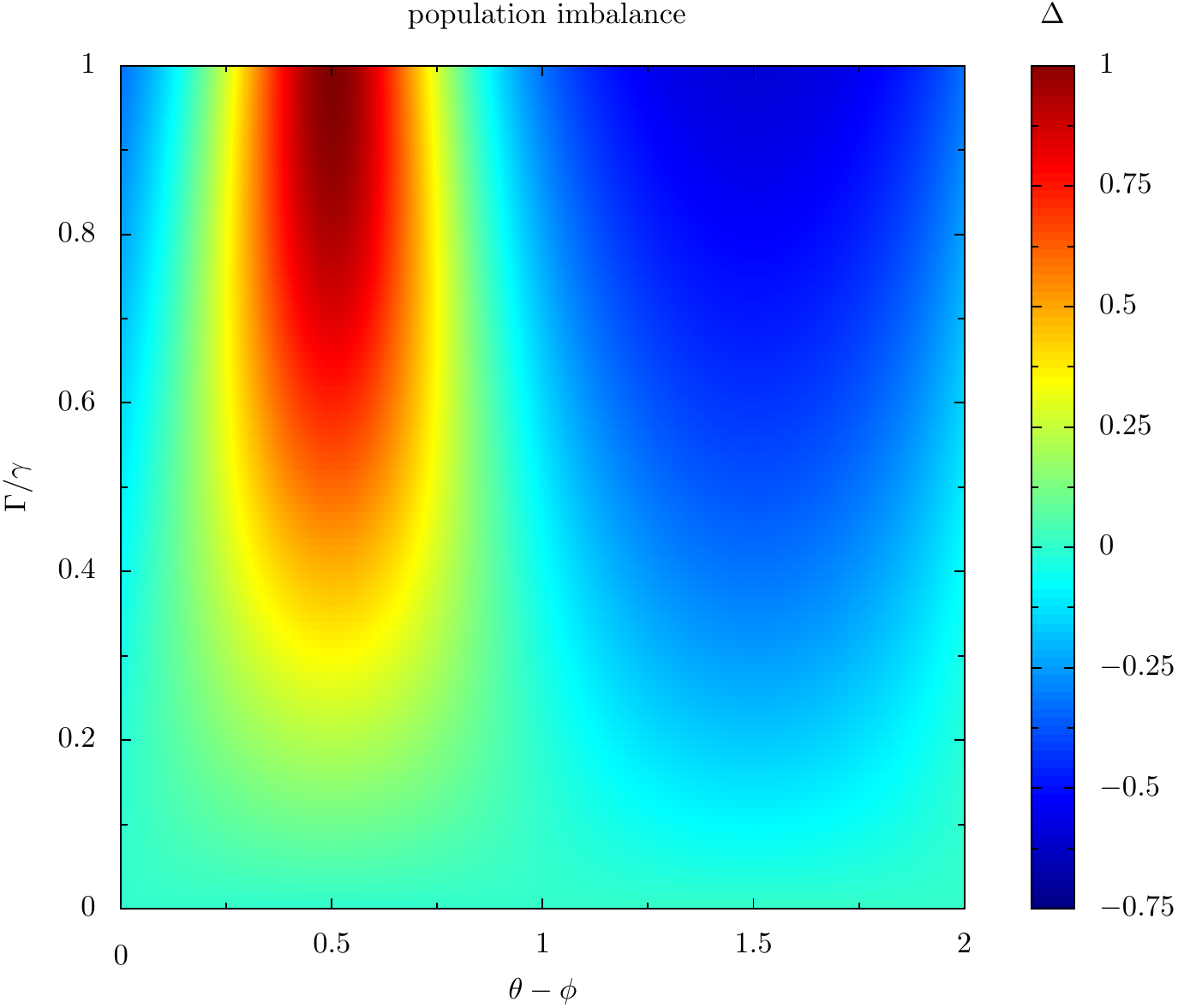}
 \caption{ \textbf{Phase-dependent population imbalance of coupled driven-dissipative resonators.} The population imbalance $\Delta$ of the pair in the steady state [cf. Eq.~\eqref{eqapsfsdfsfap:dafssdfdsaysdshr}], as a function of the relative phase $\theta - \phi$ between the coherent and dissipative couplings, and the magnitude $\Gamma$ of the dissipative coupling (in units of the loss rate $\gamma$). In the figure, the magnitude of coherent coupling $g = \gamma/2$.}
 \label{eddy}
\end{figure}

The dynamic populations are given by the compact analytical expressions~\cite{SuppInfo}
\begin{widetext}
\begin{align}
\label{eqapp:dssdfdfdfsfbcvbcayhr}
 \langle b_1^{\dagger} b_1 \rangle =& \left( \frac{2\Omega}{\gamma} \right)^2 - 2 \left( \frac{2\Omega}{\gamma} \right)^2 \mathrm{e}^{- \frac{\gamma t}{2}} 
 + \biggl\{  1  + \left( \frac{2\Omega}{\gamma} \right)^2 \biggl\} \mathrm{e}^{- \gamma t}, \\
 \langle b_2^{\dagger} b_2 \rangle =& \left( \frac{4 \Gamma \Omega}{\gamma} \right)^2 - \left( \frac{4\Gamma \Omega}{\gamma} \right)^2 \left( 2 + \gamma t \right) \mathrm{e}^{- \frac{\gamma t}{2}} 
 + \left( \frac{\Gamma}{\gamma} \right)^2 \biggl\{  \left( \gamma t \right)^2  +  \left( 2 + \gamma t \right)^2 \left( \frac{2\Omega}{\gamma} \right)^2 \biggl\} \mathrm{e}^{- \gamma t}, \label{eqapp:dssddfdfaadfdfdfsfbcvbcayhr}
\end{align}
\end{widetext}
where Eq.~\eqref{eqapp:dssdfdfdfsfbcvbcayhr} is exactly the form for a solitary driven resonator, as if the second resonator was not there, due to the completely supressed backaction. We plot the expressions of Eq.~\eqref{eqapp:dssdfdfdfsfbcvbcayhr} and Eq.~\eqref{eqapp:dssddfdfaadfdfdfsfbcvbcayhr} in Fig.~\ref{poppp}~(i) for the case of the rather strong coupling $\Gamma = (4/5)\gamma$. This coupling arrangement allows for the second resonator population (green line) to become dominant after only a short timescale $t \sim 1/\gamma$, which is maintained through to the steady state and thus evermore.
\\


\noindent \textbf{Asymmetric coupling}\\
In general, the coupling encompassed by the model of Eq.~\eqref{eqapp:massdsdsdter} is asymmetric - with the preceding unidirectional case being the most extreme example. Most generally then (when $g \ne 0$ and $\Gamma \ne 0$), the resonator steady states become dependent on the relative phase $\theta -\phi$ as follows~\cite{SuppInfo}
\begin{widetext}
\begin{equation}
\label{eqapp:dsdsfdfxvcjhdfhdayhr}
\lim_{t \to \infty} \langle b_1^{\dagger} b_1 \rangle = \frac{ \left( 2 \gamma \Omega \right)^2 }{ 16 g^4 + 8 g^2 \gamma^2 + \left( \gamma^2 - \Gamma^2 \right)^2 + 8 g^2 \Gamma^2 \cos \left( 2 \left[ \theta - \phi \right] \right) },
\end{equation}
\begin{equation}
\label{eqapdasdp:dsfdfdssfsaysdshr}
\lim_{t \to \infty} \langle b_2^{\dagger} b_2 \rangle = \frac{ 4 \Omega^2 \left( 4 g^2 + \Gamma^2 + 4 g \Gamma \sin \left[ \theta - \phi \right] \right)^2 }{ 16 g^4 + 8 g^2 \gamma^2 + \left( \gamma^2 - \Gamma^2 \right)^2 + 8 g^2 \Gamma^2 \cos \left( 2 \left[ \theta - \phi \right] \right) },
\end{equation}
from which the analogous results in other, more specialized coupling regimes may be derived. The population imbalance measure [cf. Eq.~\eqref{eqapp:dssdfdfdfdfdsaysdshr}] is given by the rich expression
\begin{equation}
\label{eqapsfsdfsfap:dafssdfdsaysdshr}
\Delta = \frac{ 2 \gamma^2 }{ 16 g^4 + 8 g^2 \gamma^2 + \left( \gamma^2 - \Gamma^2 \right)^2 + 8 g^2 \Gamma^2 \cos \left( 2 \left[ \theta - \phi \right] \right) } - 1,
\end{equation}
\end{widetext}
and is plotted in Fig.~\ref{eddy}, as a function of the relative phase $\theta -\phi$ and the dimensionless dissipative coupling strength $\Gamma/\gamma$, for the example case where the magnitude of the coherent coupling $g = \gamma/2$. Clearly, the map of Fig.~\ref{eddy} exposes the importance of the phase as the determiner of the asymmetry of the coupling, since modulating $\theta -\phi$ allows for either positive (yellow to red) or negative (cyan to blue) steady state population imbalances while holding the magnitudes of all other parameters constant. In particular, the red area around $\theta - \phi = \pi/2$ and blue region around $\theta - \phi = 3\pi/2$ are plausible from the knowledge of the unidirectional phase conditions of Eq.~\eqref{eq:Hamsdfsdfmy}.

\begin{figure}[tb]
 \includegraphics[width=1.0\linewidth]{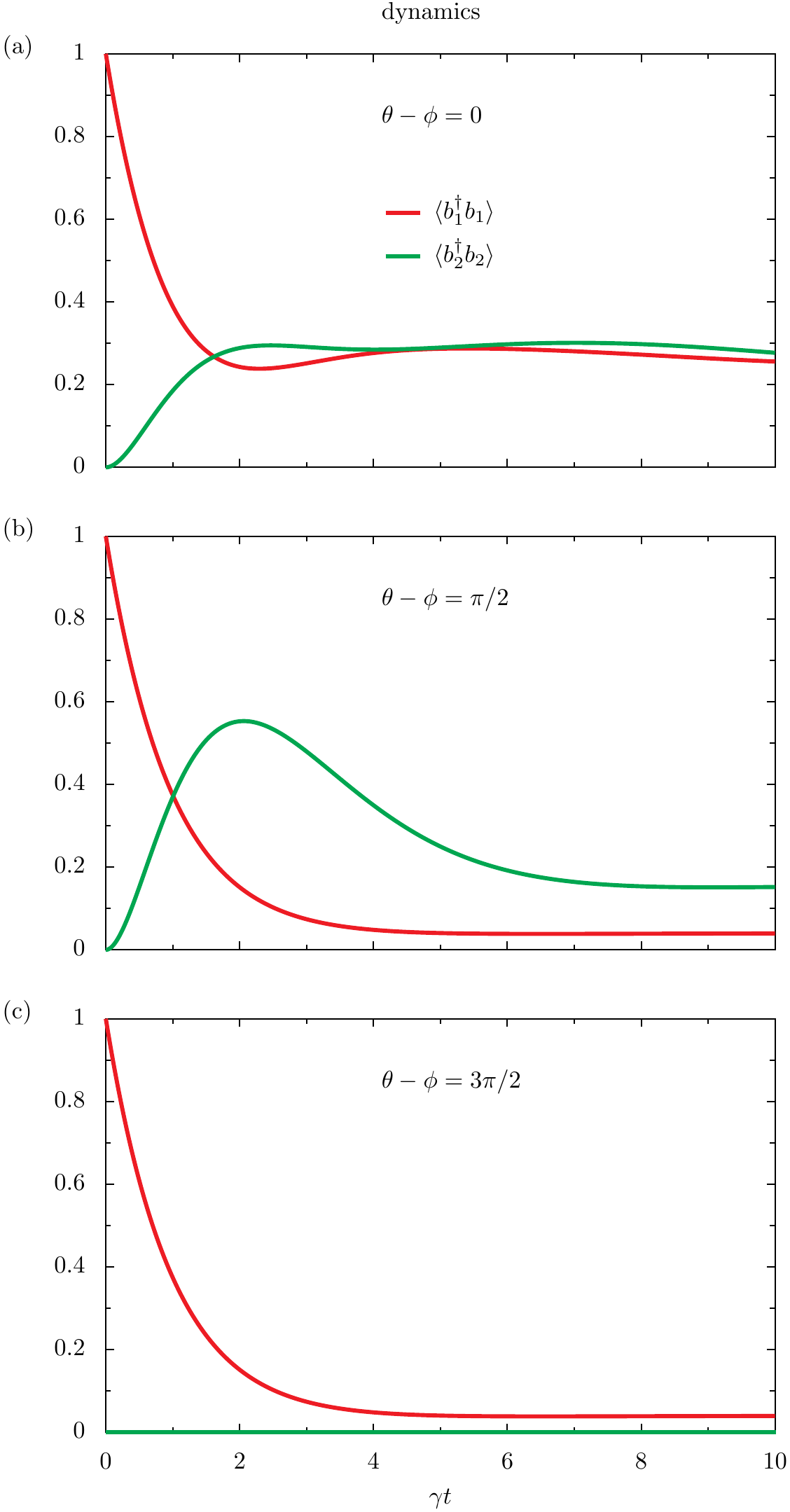}
 \caption{ \textbf{Phase-dependent population dynamics of coupled driven-dissipative resonators.} The dynamic populations $\langle b_n^{\dagger} b_n \rangle$ of the pair, as a function of time $t$ (in units of the inverse loss rate $\gamma^{-1}$). The relative phase $\theta - \phi$ between the coherent and dissipative couplings is increased from $0$ to $\pi/2$ to $3\pi/2$ upon descending the column of panels. In the figure, the first resonator is driven with an amplitude $\Omega = \gamma/10$ and the magnitudes of the coherent and dissipative couplings are $g = \gamma/2$ and $\Gamma = \gamma$, so that the magnitude unidirectional coupling condition $\Gamma = 2 g$ is met [cf. Eq.~\eqref{eq:Hamsdfsdfmy}]. }
 \label{ares}
\end{figure}

The dynamical populations are shown in Fig.~\ref{ares} for the case of $g = \gamma/2$ and maximal dissipative coupling $\Gamma = \gamma$, so that the magnitude unidirectional coupling condition is met [cf. Eq.~\eqref{eq:Hamsdfsdfmy}]. In panel (a) the relative phase is zero ($\theta - \phi = 0$) and the large incoherent-to-coherent coupling ratio of $\Gamma /g = 2$ ensures that Rabi cycles are not discernible within the damped population cycles. In panel (b) the rightwards ($\rightarrow$) unidirectional phase condition $\theta - \phi = \pi/2$  is fulfilled, such that the lack of backaction sees a purely exponential decay of the first resonator population (red line) and strong enhancement of the second resonator population (green line). Finally, in panel (c) the leftwards ($\leftarrow$) unidirectional phase condition $\theta - \phi = 3\pi/2$ is satisfied, such that the undriven second resonator is never populated, in the most dramatic realization of the directionality of the coupled system.
\\


\noindent \textbf{Conclusions}\\
We have studied a simple yet explanatorily powerful model of a pair of driven-dissipative resonators with both coherent and incoherent couplings. Our theory acts as a prototypical example of how dissipation-induced directionality may arise in quantum optical systems, with dramatic implications. In particular, we have shown how tailoring the relative magntiude and phase of the coherent and dissipative coupling can lead to highly directional and even one-way quantum transport. Our results provide perspectives for the quantum engineering of coupled resonators~\cite{Chang2018}, with applications for directional devices such as isolators, circulators and quantum batteries~\cite{Scheucher2016, Barzanjeh2017, Roushan2017}.
\\
\\


\noindent \textbf{Acknowledgments}\\
\textit{Funding}: CAD is supported by the Royal Society via a University Research Fellowship (URF\slash R1\slash 201158) and a Royal Society Research Grant (RGS\slash R1 \slash 211220). TJS acknowledges funding from the Alexander von Humboldt Foundation. \textit{Discussions}: we thank A.~I.~Fern\'{a}ndez-Dom\'{i}nguez and E.~del Valle for discussions. \textit{Data and materials availability}: There is no data in this wholly theoretical work. All necessary information is available in the manuscript and the Supplementary Information~\cite{SuppInfo}. 
\\
\\

\noindent
\textbf{ORCID}\\
C. A. Downing: \href{https://orcid.org/0000-0002-0058-9746}{0000-0002-0058-9746}.
\\
T.~J.~Sturges: \href{https://orcid.org/0000-0003-1320-2843}{0000-0003-1320-2843}.
\\
\\


\end{document}